\documentclass[aps,prapplied,reprint,superscriptaddress]{revtex4-2}

\usepackage[separate-uncertainty = true,multi-part-units=single]{siunitx}
\usepackage{hyperref}% add hypertext capabilities
\usepackage{graphicx}% Include figure files
\usepackage{float}

\begin{document}

% Use the \preprint command to place your local institutional report
% number in the upper righthand corner of the title page in preprint mode.
% Multiple \preprint commands are allowed.
% Use the 'preprintnumbers' class option to override journal defaults
% to display numbers if necessary
%\preprint{}

%Title of paper
\title{Mode Purity and Structural Analysis of X-ray Vortices Generated by Spiral Zone Plates}

\author{M. Baluktsian}
\email[]{baluktsian@is.mpg.de}
\affiliation{Max Planck Institute for Intelligent Systems, Stuttgart, Germany}
\author{L. Loetgering}
\affiliation{ACRNL, Science Park 106, 1098 XG Amsterdam, The Netherlands}
\affiliation{Vrije Universiteit, De Boelelaan 1081, 1081 HV Amsterdam, The Netherlands}
\author{G. Dogan}
\author{U. Sanli}
\affiliation{Max Planck Institute for Intelligent Systems, Stuttgart, Germany}
\author{M. Weigand}
\affiliation{HZB Bessy II, Albert-Einstein-Str. 15, 12489 Berlin, Germany }
\author{M. Rose}
\affiliation{Deutsches Elektronen-Synchrotron, Hamburg, Germany}
\author{I. Bykova}
\author{G. Sch\"utz}
\author{K. Keskinbora}
\email[]{keskinbora@is.mpg.de}
\affiliation{Max Planck Institute for Intelligent Systems, Stuttgart, Germany}

\date{\today}

\begin{abstract}
In the visible spectrum vortex beams have found various applications, ranging from optical tweezers to super-resolution imaging. Recently, these beams have been demonstrated using X-rays and electron beams. However, so far, no in-depth discussion has been carried out on the vortex quality, which could become essential for a variety of vortex applications. Here, we investigate the mode conversion efficiency (MCE), vortex structure and stability (in terms of vortex splitting) of the vortex fields generated by spiral zone plates (SZP). We have designed and fabricated SZPs with varying topological charge of both binary and kinoform profile. Kinoforms are known for their \SI{100}{\%} diffraction efficiency in the ideal case. In this work, both types are contrasted with regard to the vortex quality. Utilizing ptychographic coherent diffraction imaging and by comparing to simulations the wavefront of the generated fields is characterized. It was found, that the MCE and vortex structure exhibit the same dependencies on material and ZP properties as the diffraction efficiency (DE) and that the kinoform profile in this sense also improves the vortex quality. With growing SZP charge the MCE decreases. The results link the parameters of optics to the properties of the vortices and help to maximize the performance of ZP based vortex generators for future applications.
\end{abstract}

% insert suggested keywords - APS authors don't need to do this
%\keywords{}

%\maketitle must follow title, authors, abstract, and keywords
\maketitle

% body of paper here - Use proper section commands
% References should be done using the \cite, \ref, and \label commands
\section{Introduction}
Optical vortices have been subject to extensive research in recent years due to their attractive properties. Compared to a plane wave, whose wavefront is a plane surface, the wavefront of an optical vortex is a helical surface with an undefined phase - the so-called phase singularity - at the beam axis. Due to this internal structure an optical vortex carries orbital angular momentum (OAM) giving rise to various interesting phenomena and applications \cite{RN349,RN401}. Optical vortices are also of high interest in the short wavelength range. For instance, X-ray beams carrying OAM have been suggested for OAM dichroism at the transition metal $K$-edges \cite{RN389}, magnetic helicoidal dichroism (MHD) on magnetic surfaces \cite{Fanciulli2021,Fanciulli2022}, dichroic photoelectric effect \cite{DeNinno2020}, OAM beam-molecule interaction based resonant inelastic X-ray scattering (RIXS) spectroscopy \cite{RN391,RN392}. Further, phase sensitive imaging \cite{RN386}, initiation of optomagnetism in fullerenes \cite{RN390}, scattering of OAM X-rays from magnetic vortices \cite{RN393}, ultrafast generation of skyrmionic defects \cite{RN394}, initiation of circular dichroism in non-chiral objects by symmetry breaking \cite{RN395} and probing of molecular chirality \cite{Ye2019} have a been indicated as applications of X-ray vortices. While in the visible spectrum optical vortices can easily be generated with e.g. spatial light modulators, the generation of X-ray vortices is still an elaborate task. The first X-ray vortex was generated by Peele \textit{et al.} using a spiral staircase phase plate (SPP) \cite{RN368}. Since then a variety of extreme ultra violet (EUV) and X-ray vortex generation schemes have been proposed and realized, which can be divided into two domains: source-based and beam shaping approaches. Source-based methods have been demonstrated using FEL and synchrotron radiation as well as HHG. Hemsing \textit{et al.} introduced and utilized mechanisms such as echo enabled harmonic generation (EEHG) and high-gain high-harmonic generation (HGHG) for the generation of X-ray vortices \cite{RN369, RN370, RN371, RN372}. In another approach the second-harmonic radiation of an helical undulator has been first theoretically anticipated \cite{RN374} and later demonstrated to deliver OAM using synchrotron and FEL radiation \cite{RN375,RN373,Hemsing2020}. Further schemes have been proposed or demonstrated which are based on HHG and wave mixing arrangements, Compton and stimulated Raman backscattering or relativistic harmonics on the surface of a solid \cite{RN382,RN376,RN377,RN381,RN378,RN379,RN380}\newline
In beam shaping approaches optics such as the SPP are placed into the beam path, which imprint the spiral phase onto the incoming radiation. In contrast to source-based approaches these arrangements are technically easier to operate (for instance when switching between helicities) as extensive and often time-consuming modifications, to e.g. beam lines, are not required. Enabled through advances in the design and nanofabrication the range of X-ray singular optics has been in the last years augmented by introducing refractive SPPs \cite{Seiboth2019}, spiral photon sieves \cite{RN383}, forked gratings \cite{RN384,Lee2019} and spiral zone plates (SZPs) \cite{RN385,RN386,RN387,Ribic2017,RN345}. Vortex dipoles originating from square aperture diffraction \cite{RN396}, the rotating beams from helical beam FZPs \cite{RN397} or scattering on artificial spin ice (ASI) metasurfaces \cite{Woods2021} have also been proposed as sources of OAM X-ray beams. \newline
In this work, we focus on SZPs as vortex generators because of their versatility: the usability in a relatively wide energy and setup range (i.e. laser \cite{RN400}, synchrotron, FEL setups \cite{RN388} or even electron beams \cite{RN398}) and the simultaneous ability to focus down to spatial dimensions becoming essential for some predicted effects.\newline
So far, many schemes for generation of X-ray OAM beams have been discussed and presented. However, no evaluation of the vortex character formed by these methods, have been made. The vortex quality, though, might become crucial for some aforementioned applications particularly OAM dichroism, OAM light-molecule and -magnetic vortex interaction which arise out of the radial variation of a vortex. In this context, a kinoform ZP \cite{RN344,RN367}, which is known for its characteristic of having a single diffraction order and 100$\%$ diffraction efficiency (DE) in the lossless limit, is expected to create a well-defined vortex structure compared to the binary ZP. \newline 
However, to date only binary SZPs have been reported. In this work, we present the first kinoform SZP and compare its properties to those of binarized SZPs with respect to the vortex composition and quality. Utilizing ptychographic coherent diffraction imaging (PCDI or ptychography) at BESSYII, we measure the generated vortex fields and reconstruct the spiral optics. By performing mode and vortex structure analysis on these reconstructions the wavefront shaping capabilities are investigated dissociated from aberrations that may be caused by the beamline optics. Consequently, we are able to discuss and identify vortex structure modifications caused by ZP aberrations. Further, the impact of ZP and material properties on the vortex quality and dynamics are discussed.

\section{Zone plates and Experimental Setup}
\begin{figure*}
\includegraphics{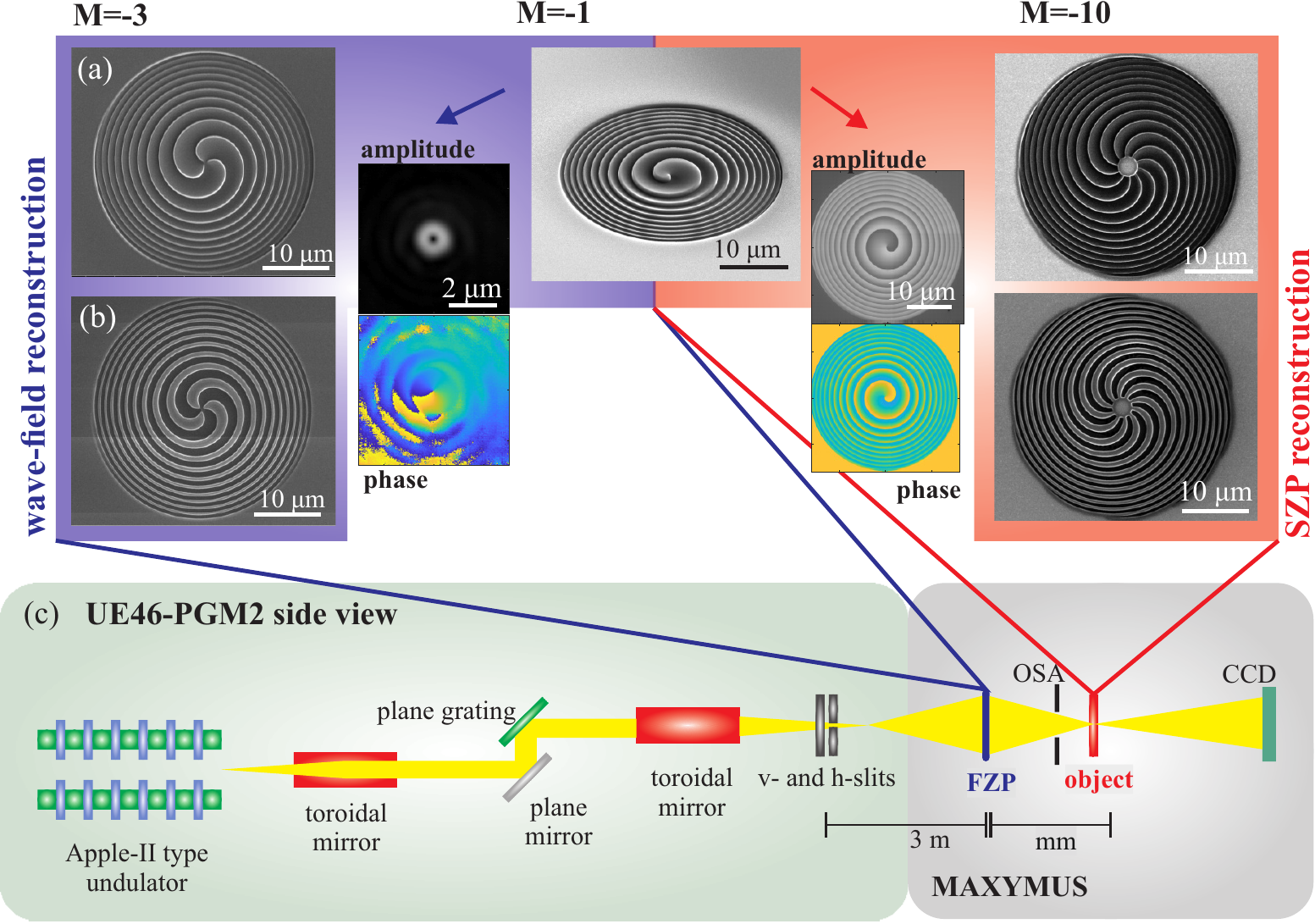}
\caption{\label{fig:01-Setup} \textbf{Fabricated SZPs and Experimental Setup.} Scanning electron microscopy (SEM) micrographs of fabricated $M=-1,-3$ and $M=-10$ (a) kinoform and  (b) binary SZPs. (c) schematic sideview of microscopy beamline UE46-PGM2 and the scanning microscope MAXYMUS at BESSYII. The endstation consists of the main components: FZP, order selecting aperture (OSA), object and charge-coupled device (CCD) detector. Two different measurements are performed. The optics are either placed at FZP or object position obtaining either wave-field reconstructions or SZP reconstructions.}
\end{figure*}

%The SZP function of topological charge $M$ is given by the product of the kinoform phase function in the paraxial approximation $\exp(-i\pi r^2/\lambda f)$ and the Hilbert phase $e^{im\phi}$ \cite{RN386}.
The projected thickness function of an SZP of topological charge $M$ is given by \cite{RN386,Paganin2006}
\begin{equation}
\label{eq:SZPfunction}
T(r,\phi) = \left(-\frac{\pi r^2}{\lambda f} + M\phi\right)_{\mathrm{mod}T_{2\pi}}
\end{equation}
where $r$, $\phi$ are the polar coordinates, $\lambda$ the wavelength and $f$ the focal length of the ZP lens. Taking modulo $T_{2\pi}$, where $T_{2\pi}$ denotes the ideal material thickness that corresponds to a $2\pi$ phase shift of the monochromatic X-ray in the material, the kinoform SZP lens shown in Fig.~\ref{fig:01-Setup}~(a) is obtained. Binarizing Eq.~\ref{eq:SZPfunction} yields the binary SZP shown in Fig.~\ref{fig:01-Setup}~(b). SZPs with various topological charges were fabricated on a \SI{1}{\mu m}-thick Si$_3$N$_4$ membrane using an extended version of the direct-write ion beam lithography approach reported in \cite{RN344} (see the Supplementary Material). The ZPs are designed with a focal length of $f=$ \SI{12.4}{mm} at \SI{1200}{eV}, an outermost zone period of $\Lambda=$ \SI{800}{nm} and a diameter of $\o$=\SI{32}{\mu m}. The ZP thickness are determined graphically from the SEM image and are measured to be \SI{823\pm12}{nm} for the $M=1$ kinoform and \SI{701\pm47}{nm} for the binary and \SI{766\pm17}{nm} and kinoform $M=3,10$ SZPs.\newline 
The fabricated optics are characterized in a scanning transmission x-ray microscope MAXYMUS (magnetic x-ray microscope with ultrahigh vacuum spectroscopy) at the UE46-PGM2 beam line of BESSY II (Helmholtz Zentrum Berlin). Figure~\ref{fig:01-Setup}~(c) depicts a schematic of the beamline and the microscope. A partially coherent beam with a spectral range of \SIrange{200}{1900}{eV} and tunable polarization is focused on a crossed pair of exit slits. The beamline has a resolving power of $E/ \Delta E \sim 10^3$ \cite{RN347}. The FZP is located \SI{3}{m} downstream of the exit slits focusing the beam onto the object. An order-selecting aperture ($\o=$\SI{15}{\mu m}) filters out unwanted higher diffraction orders of the FZP. The object is mounted onto a interferometer controlled piezoelectric stage enabling $xy-$translation with a precision of \SI{2.5}{nm}. An X-ray charge coupled device detector (256$\times$256 pixels of \SI{48}{\mu m} pixel size) being located \SIrange{9}{16}{cm} downstream of the specimen is used to record at each sample position the diffraction patterns resulting from the interaction of the vortex beam with the sample. The collection of diffraction patterns (spectogram) forms the dataset for PCDI (for details on the ptychographic reconstruction algorithm see \cite{RN397}). For the characterization of the SZPs two different measurements are performed. In the first the SZPs are placed at the FZP position and their focal plane wave fields are reconstructed. In the second complex-valued reconstructions of the SZPs are obtained by placing the SZPs at the object plane (cf. Fig.~\ref{fig:01-Setup}~(c)).
%info on diffraction efficiency

\section{Mode purity}
\begin{figure}
\includegraphics{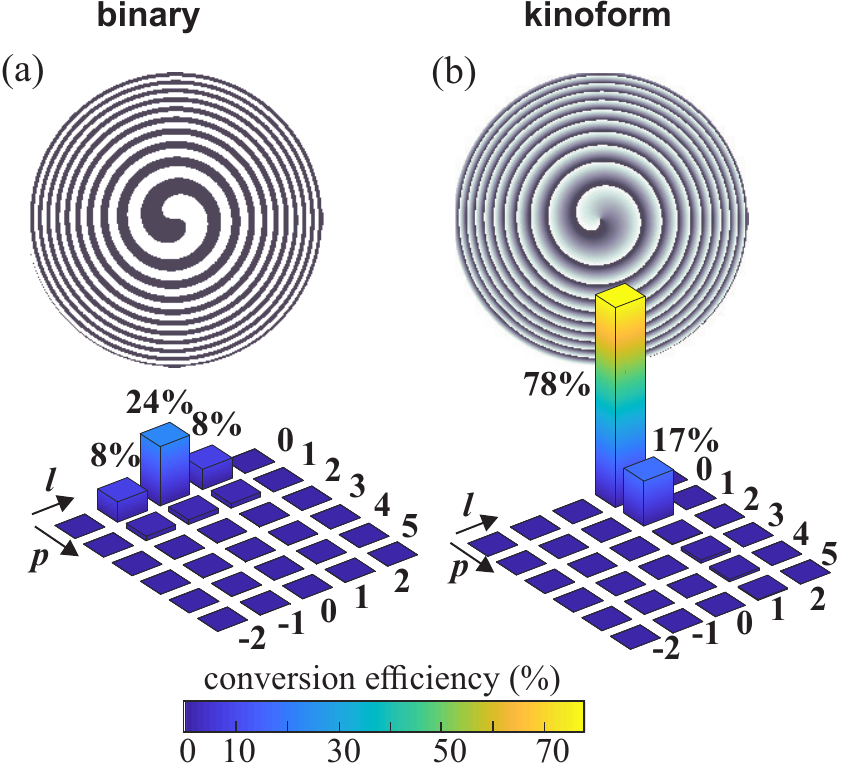}
\caption{\label{fig:02-MCEtheory} \textbf{Mode analysis of simulated ideal $M=1$ binary and kinoform SZP.} Bar graph representations of the conversion efficiencies of a $M=1$ (a) binary and (b) kinoform SZP for a $(0,0)$ input mode. The kinoform converts with a ten times larger efficiency into the fundamental mode than the binary and has no unconverted mode contribution.}
\end{figure}

\begin{figure*}
\includegraphics{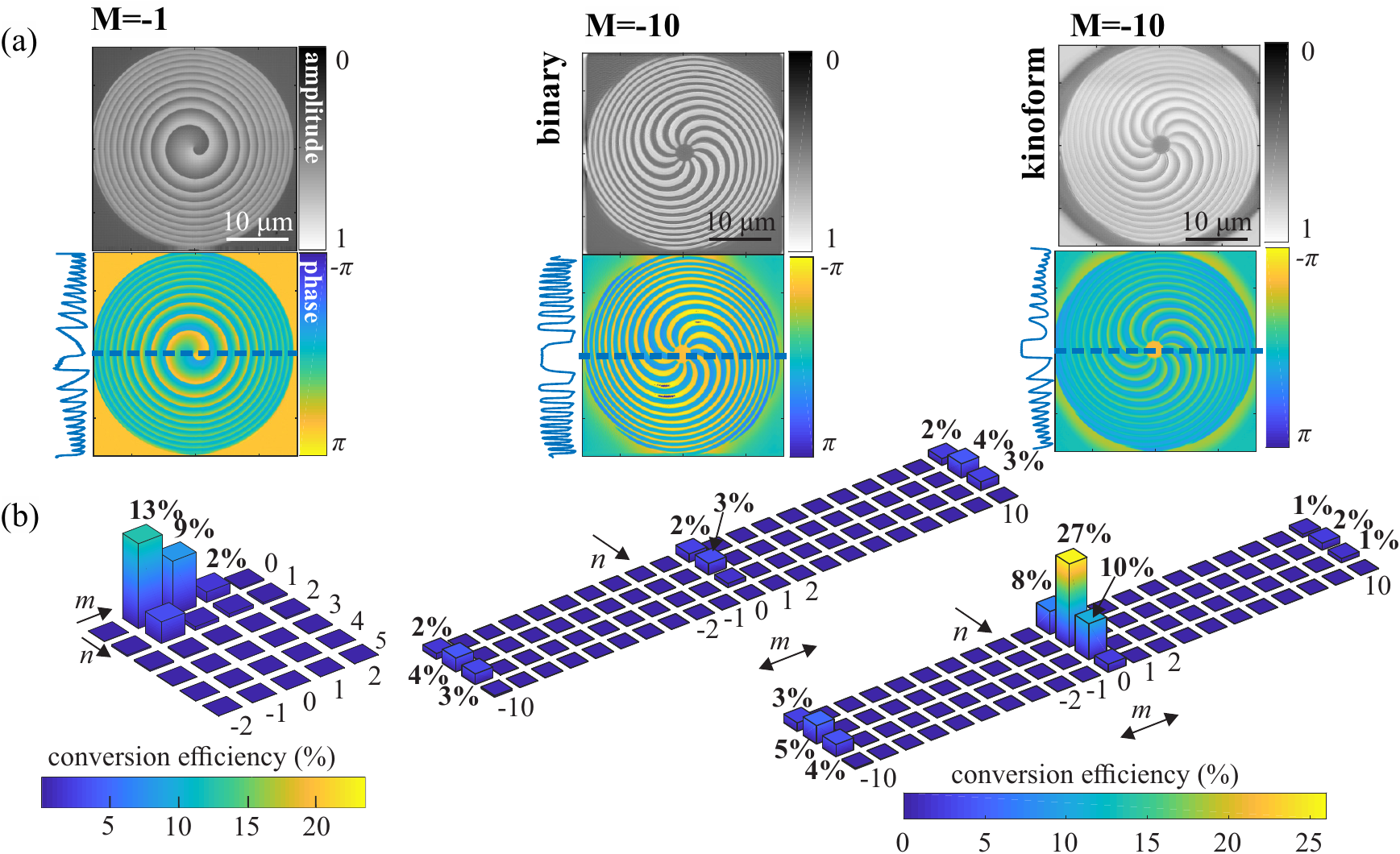}
\caption{\label{fig:03-ExpResults1} \textbf{Experimental Results of $M=-1$ and $M=-10$ SZPs.} (b) Ptychographically retrieved SZP reconstructions of the $M=-1$ kinoform at \SI{700}{eV} and $M=-10$ SZPs at $\approx$ \SI{685}{eV} with the corresponding cross sections. (b) From reconstructions calculated conversion efficiencies for a $(0,0)$ input mode.}
\end{figure*}

\begin{figure}
\includegraphics{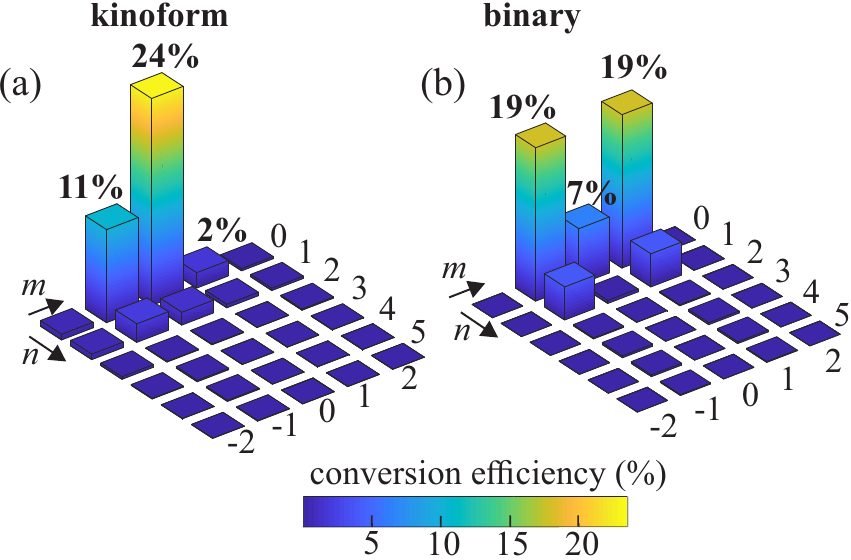}
\caption{\label{fig:04-MCEsim1} \textbf{Conversion efficiencies of $M=1$ SZPs simulated with real parameters.} Calculated relative intensities of a $(0,0)$ input mode modified by simulated $M=1$ (a) kinoform and (b) binary considering phase shift and absorption in silicon nitride at \SI{700}{eV} and \SI{823}{nm} ZP thickness.}
\end{figure}

The SZP is a mode converter, i.e. it up- and downconverts the helicity of the incoming beam. The mode distribution of a beam converted by an SZP converted beam can be investigated by calculating the relative intensities \cite{RN365}
\begin{equation}
\label{eq:MCE}
I_{m'n',mn}=|\langle u^{LG}_{m'n'}| \exp({iT(r,\phi)}) |  u^{LG}_{mn}\rangle|^2
\end{equation}
of an LG mode with indices $(m,n)$ into an LG mode with indices $(m',n')$ (see the Supplemental Material). Similar to an SPP \cite{RN365} the SZP is not a pure mode converter. Fig.~\ref{fig:02-MCEtheory} shows the mode contents of a $(0,0)$ mode modified by a (pure phase shifting) kinoform and binary (alternating transmissive and fully opaque zones) SZP phase function defined by Eq.~\ref{eq:SZPfunction}.
Note, that the topological charge $M$ of the SZP dictates the azimuthal mode number $m'$ and therefore the mode ($m'=M, n'=0$) is referred to as the fundamental mode. The kinoform converts with \SI{78}{\%} efficiency into the fundamental mode being ten times larger than the fundamental mode of the binary. The residual \SI{22}{\%} are distributed over higher radial modes. In contrast to the kinoform, the binary exhibits both polarities and a significant portion of the unconverted incoming mode (\SI{23}{\%}). The mode with the opposite polarity of the fundamental mode is assigned to the virtual focus of the binary being located upstream of the SZP and diverging downstream of the SZP. In contrast, the kinoform has (in the ideal case with no absorption within the optic) no virtual focus, which is, as the single diffraction order and \SI{100}{\%} DE, a characteristic of the kinoform. Note, that in the experiment, this opposite fundamental mode will be filtered out by the OSA. \newline %\subsection{Experimental Results}
The mode conversion properties of the fabricated SZPs are studied by calculating Eq.~\ref{eq:MCE} using the ptychographically obtained complex valued reconstructions of the $M=1$ and $M=10$ SZPs shown in Fig.~\ref{fig:03-ExpResults1}~(a).
Their calculated relative intensities are depicted in bar graph representation in Fig.~\ref{fig:03-ExpResults1}~(b).\newline 
The real $M=1$ SZPs exhibits a 7 times lower intensity into the fundamental mode than expected. Further, an unconverted mode and opposite fundamental mode contribution emerges.\newline
The fundamental mode contribution of the $M=10$ charged kinoform SZP drops significantly achieving with \SI{3}{\%} only a fourth of the single-charged contribution. The fundamental mode of the binary $M=10$ SZP resembles the kinoform. As expected, the binary has symmetrical fundamental and opposite fundamental azimuthal mode index contributions while the kinoforms fundamental mode is slightly larger than its opposite polarity. In contrast to Fig.~\ref{fig:02-MCEtheory}~(b) the azimuthal mode index of the fundamental mode is reversed. This results from the negative sign in the exponent of the phase shift term expression of a wave propagating in a medium.%\subsection{Discussion}
The deviation of the measured $M=1$ kinoform SZP from the predicted values results from absorption and phase shift within the silicon nitride. In Fig.~\ref{fig:04-MCEsim1} the observed relative intensities of the $M=1$ kinoform are reproduced in a simulation in which a generalized SZP function is used.
Further, the $M=1$ kinoform is compared to a $M=1$ binary. The generalized SZP function considers phase retardation and attenuation of the propagating wave within the optic at \SI{700}{eV}. Note, that the kinoform and binary in Fig.~\ref{fig:02-MCEtheory} are special cases of this function (see Supplemental Material).\newline 
Fig.~\ref{fig:05-MCEsim2} shows that the simulated SZP kinoform has as well a 7 times lower conversion efficiency into the fundamental mode and an unconverted mode contribution compared to Fig.~\ref{fig:02-MCEtheory}~(b). The simulated binary SZP exhibits a two times larger MCE into the fundamental mode than the kinoform. The superior MCE of the binary is similar to its DE, where at 700 eV and 823 nm ZP thickness the binary has a larger DE than the kinoform because the binary is closer to optimum ZP thickness for maximum DE \cite{RN348}. Indeed, the MCE has the same optimum ZP thicknesses as the DE.\newline
\begin{figure}
\includegraphics{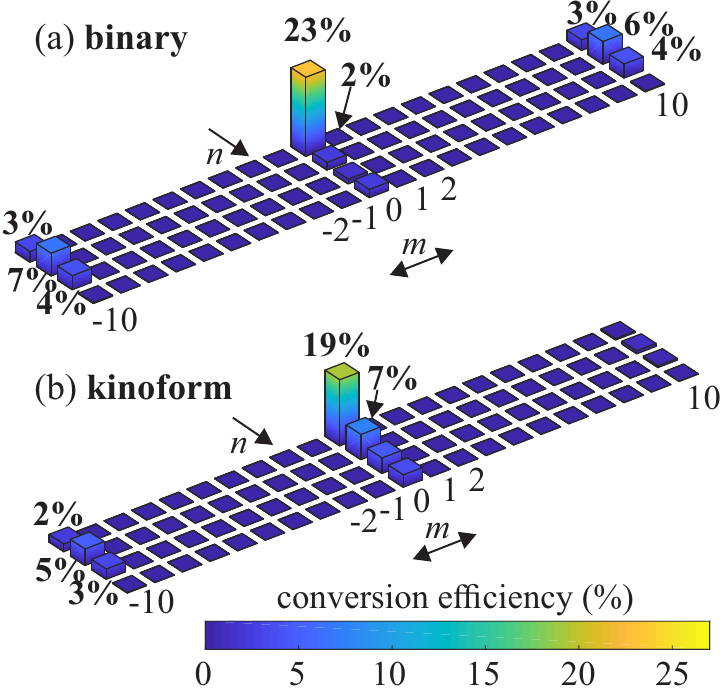}
\caption{\label{fig:05-MCEsim2} \textbf{Conversion efficiencies of simulated $M=10$ SZPs.} Calculated relative intensities of a $(0,0)$ input mode modified by simulated $M=1$ (a) \SI{701}{nm} thick kinoform and (b) \SI{766}{nm} thick binary considering phase shift and absorption in silicon nitride at \SI{685}{eV}.}
\end{figure}
Fig.~\ref{fig:05-MCEsim2} depicts the relative intensities for the simulated $M=10$ SZPs taking into account the 3 um wide unpatterned central area of the fabricated SZPs. The simulated and measured fundamental mode contributions are in good qualitative agreement. With comparative simulations it can be shown that the deviation of the ZP thickness from optimum and the unpatterned central area strongly increase the unconverted mode contribution. This is in contradiction to the measured binary SZP where a strong unconverted mode is expected due to the unpatterned central area.\newline
\newpage
\section{Vortex structure}
\begin{figure*}
\includegraphics{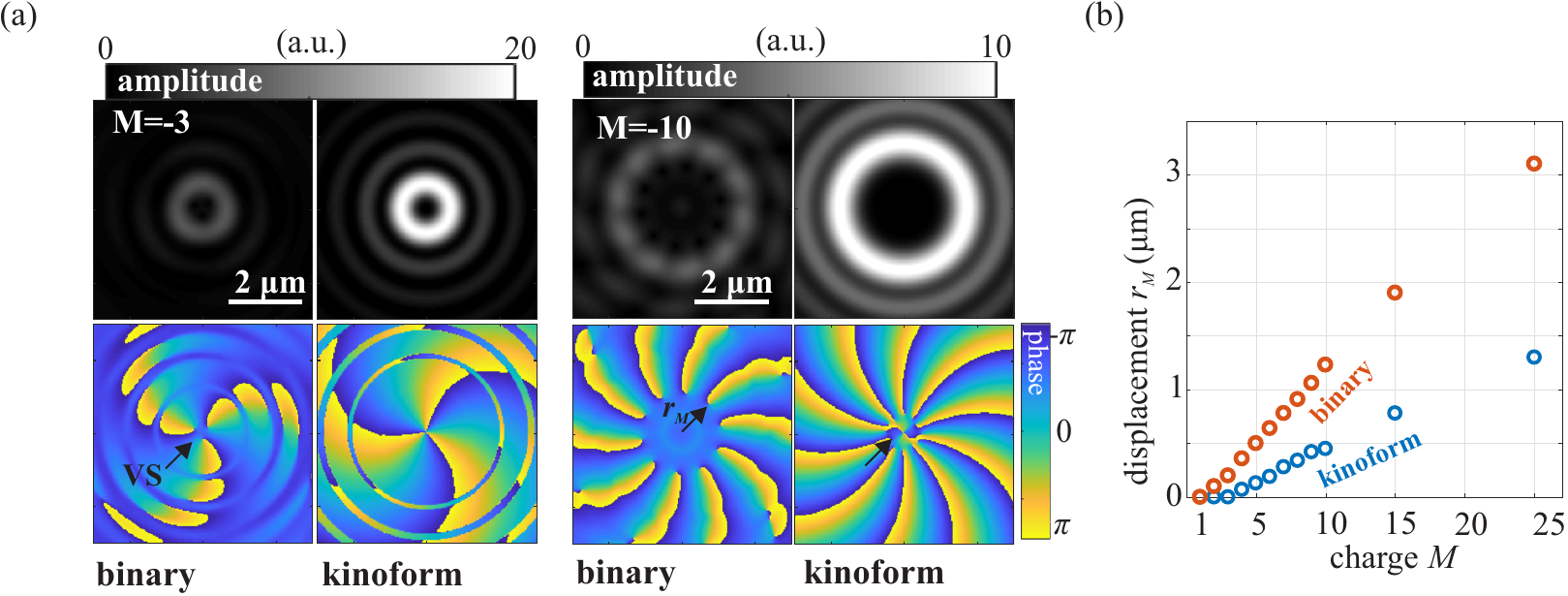}
\caption{\label{fig:06-VDtheory} \textbf{Vortex splitting in dependence of SZP charge $M$.} (a) Focal plane wave field of $M=1$ and $M=10$ binary and kinoform SZPs. (b) Radial displacement $r_l$ of the $m=l$ in dependence of SZP charge $M$. The vortex splitting increases with the SZP charge.}
\end{figure*}

\begin{figure*}
\includegraphics{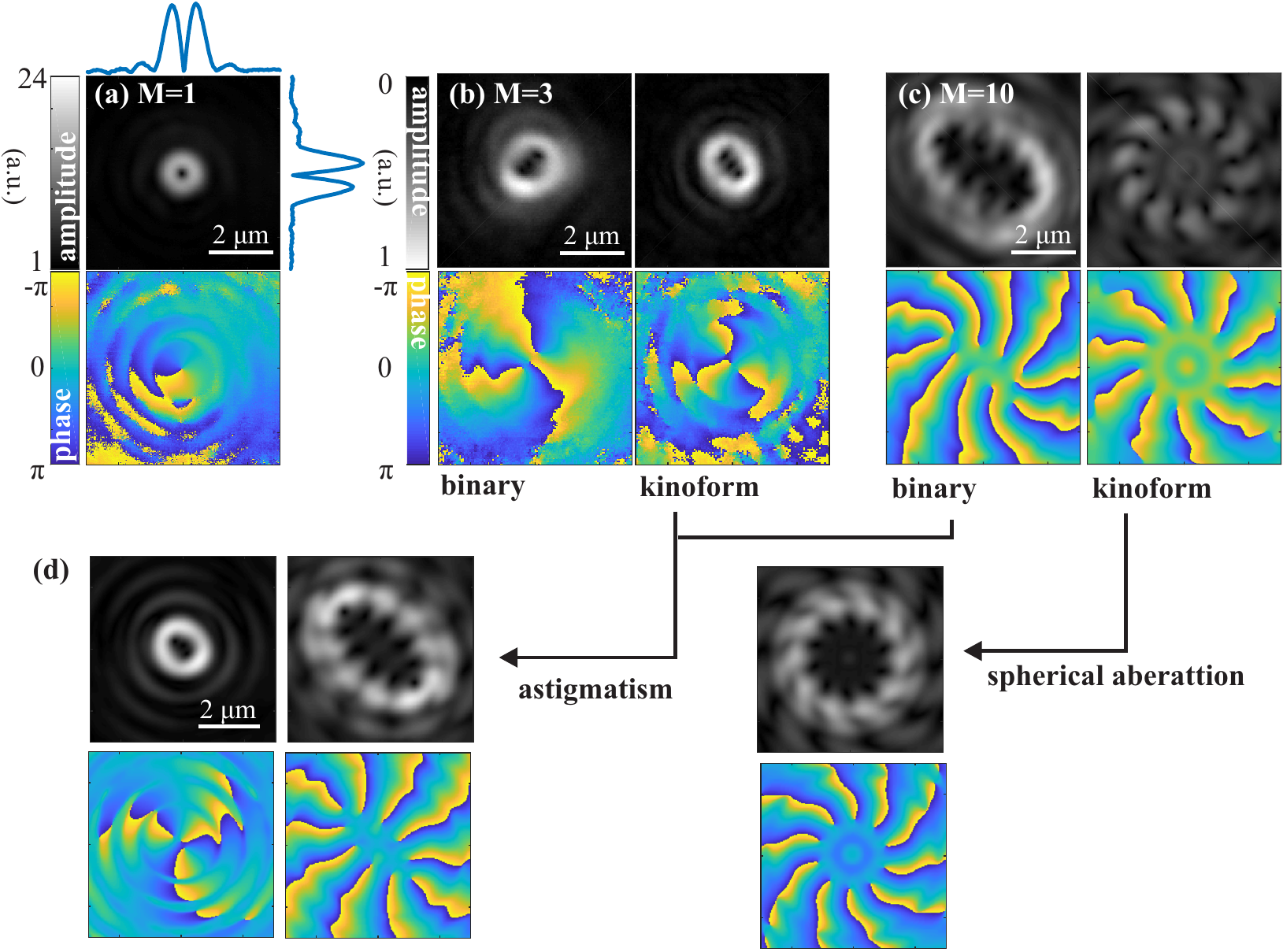}
\caption{\label{fig:07-VDmeas} \textbf{Ptychographic reconstructions and simulations of focal plane wave fields.} Focal plane wave fields of the (a) $M=-1$ (\SI{800}{eV}) and (b) $M=-3$ \SI{600}{eV} SZPs. (c) Numerically propagated focal plane wave fields of the measured $M=-10$ (\SI{685}{eV}) SZP function from Fig.~\ref{fig:03-ExpResults1}. In the experiment additional distortions of the vortex field are observed. (d) Wave fields of simulated SZP functions $M=-3$ \SI{766}{nm} ZP thickness, $M=-10$ \SI{701}{nm} thick binary and \SI{766}{nm} thick kinoform at \SI{685}{eV}. }
\end{figure*}

In Fig.~\ref{fig:06-VDtheory} the numerically propagated focal plane wave fields of binary and kinoform SZPs defined by Eq.~\ref{eq:SZPfunction} are compared for topological charges $M=-3$ and $M=-10$.
The full widths at half maximum (FWHM) defined by $\omega_{x,y}=2\sqrt(2\ln(2))\sigma_{x,y}$ with $\sigma_{x,y}$ being the standard deviation in x- and y-direction of the wave fields are \SI{1.1}{\mu m}($M=1$ not shown here), \SI{2.1}{\mu m} ($|M|=3$) and \SI{5.4}{\mu m} ($|M|=10$). For topological charges $|M|>1$ the higher-order vortices break down into a number of first-order vortices being equal to the absolute topological charge of the higher-order vortex \cite{Paganin2006,Gbur2016}. As depicted in Fig.~\ref{fig:06-VDtheory}~(a) and (b) the vortex splitting increases with increasing topological charge, while the amplitude significantly decreases. The kinoform exhibits in this case a brighter and more homogeneous intensity distribution, which is an immediate result of the superior DE compared to the binary.\newline %\subsection{Experimental Result}
Using ptychography, the focal plane wave fields of the $M=-1$ kinoform and $M=-3$ SZPs (cf. Fig~\ref{fig:01-Setup}~(a)) were measured at \SI{800}{eV} and \SI{600}{eV}, respectively. The results are presented in Fig.~\ref{fig:07-VDmeas}~(a)-(b).
Further, next to the measurements the numerically propagated wave fields of the $M=-10$ SZP reconstructions from Fig.~\ref{fig:03-ExpResults1} are depicted.\newline
As expected vortex splitting is observed for $|M|>1$. The radial displacement of the single-charged vortices for $M=-10$ kinoform is nearly 3 times larger compared to Fig.~\ref{fig:06-VDtheory}. The $M=-1$ vortex exhibits a slightly asymmetric amplitude distribution with varying beam widths $\omega_x=$\SI{1.52}{\mu m} and $\omega_y=$\SI{1.39}{\mu m}. Further, additional vortex distortions are observed.\newline  %\subsection{Discussion}
Focal plane wave fields of simulated general SZP functions, that equal in ZP properties to the measured SZPs, are calculated for comparison and shown for $M=-3$ and $M=-10$ in Fig.~\ref{fig:07-VDmeas}~(d).
%\begin{figure*}
%\includegraphics{08-VDsim}
%\caption{\label{fig:08-VDsim} \textbf{Focal plane wave fields of simulated SZPs functions with aberrations.}(a) $M=1$ kinoform SZP function with phase shift and absorption at \SI{800}{eV} and \SI{823}{nm} ZP thickness. (b) $M=3$ SZP function at \SI{600}{eV}, \SI{701}{nm} (binary) and \SI{766}{nm} (kinoform) ZP thicknesses and astigmatism. (c) $M=10$ SZP function at \SI{685}{eV}, same ZP thicknesses as the $M=3$ SZPs and with astigmatism and spherical aberration, respectively. }
%\end{figure*}
The simulated $M=-1$ vortex field (not shown here) exhibits a to the experiment similar asymmetric amplitude distribution with beam widths $\omega_x=$\SI{1.64}{\mu m} and $\omega_y=$\SI{1.59}{\mu m}. The observed vortex distortions are reproduced with Seidel aberrations added to the wave field. It is found that the vortex distortions are caused by astigmatism resulting in an anisotropic elliptically shaped vortex field and spherical aberration manifesting as 'smearing' of the amplitude distribution.\newline
Given the discussion on mode analysis above the MCE crucially depends on the material properties and is maximized at the optimum ZP thickness for maximum DE. The SZP generates a superposition of modes, which as discussed in \cite{RN354} leads to a perturbation of the fundamental mode giving rise to a displacement of the singularity and a vortex splitting of higher-order vortex fields given by the following expression \cite{RN354}
\begin{equation}
r_m = w_0 \left(\frac{E_G}{|u_{m0}|}\right)^{1/|m|}\frac{z^2+k^2w_0^4/4}{k^2w_0^2/2},
\end{equation} 
where $E_G$ denotes the small amplitude of the Gaussian beam $u_{00}$ that perturbs the vortex field, $k$ the wave vector and $w_0$ the Gaussian beam waist at $z=0$. Based on this expression an SZP with a maximized DE (and MCE) exhibits due to low contributions of non-fundamental modes the lowest vortex splitting. This leads to the following conclusions: (1) Not only the mode purity but also the vortex structure is best at the optimum ZP thickness for maximum DE. (2) A modified ZP design, that generates a superposition of less modes that perturb the fundamental mode, can be employed to minimize vortex splitting. This has been demonstrated in \cite{RN353} with a first-order compensation method.\newline

\section{Conclusions}
In conclusion, we have fabricated SZPs with various topological charges and characterized their wave front by means of ptychography. Wave front characterization was performed by computational reconstruction of both the beams and the SZPs. Mode decomposition analysis was applied to the SZP reconstructions observing a conversion efficiency of \SI{13}{\%} for the $|M|=1$ kinoform into the fundamental mode, which decreases with increasing topological charge achieving \SIrange{2}{3}{\%} for $|M|=10$ SZPs. The measurements were compared to simulations observing similar dependencies on ZP and material properties as is known from the diffraction efficiency (e.g. kinoform vs. binary profile). The reconstructed fields of $M>1$ SZPs showed a vortex split as predicted by theory of vortex dynamics \cite{Paganin2006}. Additional distortions observed in the vortex structure of the measured $M=-3$ and $M=-10$ beams were by means of simulations identified to originate from spherical aberration and astigmatism. Finally, the relation of vortex splitting to conversion efficiency was discussed concluding that the lowest vortex splitting is expected at maximum conversion efficiency. \newline
In this work, it was shown that the vortex quality with regard to mode purity and vortex splitting is the highest when the diffraction efficiency is maximized. These findings help to optimize the SZP for applications where the results are sensitive to the discussed vortex properties \cite{RN389,RN391,RN392}. Regarding this, an interesting future route will be the investigation of modified ZP designs (similar to the approach reported in \cite{RN353}) to further improve mode purity and minimize vortex splitting.

\section{Acknoledgements}
We thank HZB for the allocation of synchrotron radiation
beamtime.

\bibliography{pub_SZP}

\end{document}

% Use the \preprint command to place your local institutional report
% number in the upper righthand corner of the title page in preprint mode.
% Multiple \preprint commands are allowed.
% Use the 'preprintnumbers' class option to override journal defaults
% to display numbers if necessary
%\preprint{}

%Title of paper
\title{Supplemental Material for \\
Mode Purity and Structure Analysis of Optical X-ray Vortices Generated by Spiral Zone Plates}

% repeat the \author .. \affiliation  etc. as needed
% \email, \thanks, \homepage, \altaffiliation all apply to the current
% author. Explanatory text should go in the []'s, actual e-mail
% address or url should go in the {}'s for \email and \homepage.
% Please use the appropriate macro foreach each type of information

% \affiliation command applies to all authors since the last
% \affiliation command. The \affiliation command should follow the
% other information
% \affiliation can be followed by \email, \homepage, \thanks as well.
%\author{}
%\email[]{Your e-mail address}
%\homepage[]{Your web page}
%\thanks{}
%\altaffiliation{}
%\affiliation{}

\author{M. Baluktsian}
\email[]{baluktsian@is.mpg.de}
\affiliation{Max Planck Institute for Intelligent Systems, Stuttgart, Germany}
\author{L. Loetgering}
\affiliation{ACRNL, Science Park 106, 1098 XG Amsterdam, The Netherlands}
\affiliation{Vrije Universiteit, De Boelelaan 1081, 1081 HV Amsterdam, The Netherlands}
\author{G. Dogan}
\author{U. Sanli}
\affiliation{Max Planck Institute for Intelligent Systems, Stuttgart, Germany}
\author{M. Weigand}
\affiliation{HZB Bessy II, Albert-Einstein-Str. 15, 12489 Berlin, Germany }
\author{M. Rose}
\affiliation{Deutsches Elektronen-Synchrotron, Hamburg, Germany}
\author{I. Bykova}
\author{G. Sch\"utz}
\author{K. Keskinbora}
\email[]{keskinbora@is.mpg.de}
\affiliation{Max Planck Institute for Intelligent Systems, Stuttgart, Germany}
%Collaboration name if desired (requires use of superscriptaddress
%option in \documentclass). \noaffiliation is required (may also be
%used with the \author command).
%\collaboration can be followed by \email, \homepage, \thanks as well.
%\collaboration{}
%\noaffiliation

\date{\today}

%\begin{abstract}
% insert abstract here
%\end{abstract}

% insert suggested keywords - APS authors don't need to do this
%\keywords{}

%\maketitle must follow title, authors, abstract, and keywords
\maketitle

% body of paper here - Use proper section commands
% References should be done using the \cite, \ref, and \label commands

% Put \label in argument of \section for cross-referencing
%\section{\label{}}
\section{SZP Nanofabrication}
The ZPs are patterned on a \SI{1}{\mu m}-thick Si$_3$N$_4$ membrane (Silson Ltd., United Kingdom), which are priorly coated with a \SI{10}{nm} gold layer to avoid ion beam shifts caused by membrane charging. Other than in \cite{RN344} the SZP pattern is disassembled in points arranged on a 1200$\times$1200 grid with \SI{30}{nm} step size. Each point has the physical dimension of the ion beam size and is a assigned an individual ion dose factor between 0 and 1 representing the multiplier of the user defined dwell time. A \SI{30}{keV} focused Ga-ion beam (Nova NanoLab 600, FEI) of \SI{300}{pA} ion beam current resulting in a nominal ion beam size of \SI{31}{nm} and an overlap of \SI{3.3}{\%} is used for patterning. The SZP pattern is written in \num{60} cycles with a dot dwell time of \SI{200}{\mu s} per cycle. This results in a dot dosage of\SI{0.06}{pC} per cycle and a total patterning time of $\approx$\SI{2}{h}. Image cross correlation based drift correction is applied between the cycles.\newline

\section{Mode decomposition}
\begin{figure}
\includegraphics{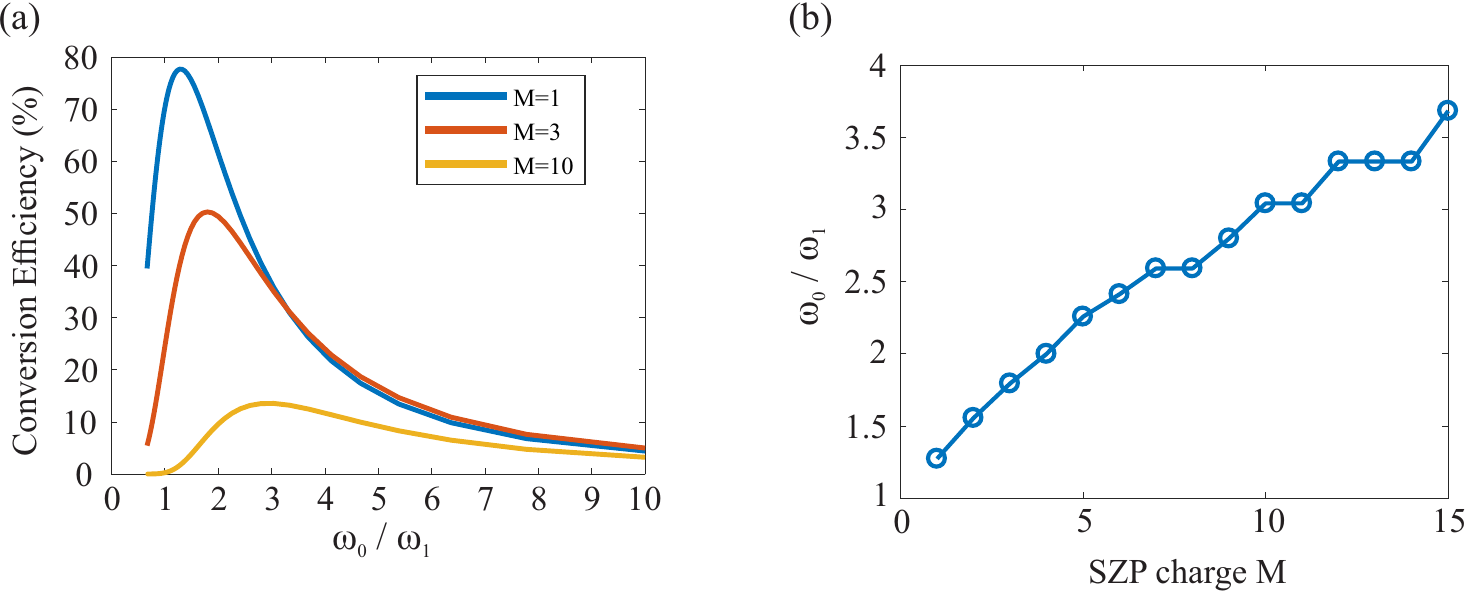}
\caption{\label{fig:S1-MCEdecomposition} \textbf{Fundamental mode MCE in dependence of beam waist ratio.} (a) Conversion efficiency $I_{M0,00}$ (fundamental mode) for a $M=1,3,10$ SZP as a function of the input beam size $\omega_0$ normalized to the output beam size $\omega_1$. The $M=1$ curve has a maximum at an optimal beam waist ratio of $\gamma = 1.3$. (b) Optimal beam waist ratio as a function of topological SZP charge $M$.}
\end{figure}

Similar to \cite{RN365} the conversion efficiency of an SZP given by
\begin{equation}
\label{eq:SZPfunction}
T(r,\phi) = \left(-\frac{\pi r^2}{\lambda f} + M\phi\right)_{\mathrm{mod}T2\pi}
\end{equation}
from a Gaussian $(0,0)$ mode into LG modes $(m',n')$ is obtained by calculating
\begin{equation}
\label{eq:MCE}
I_{m'n',00}=|\langle u^{LG}_{m'n'}(\omega_1)| \exp({iT(r,\phi)}) |  u^{LG}_{00}(\omega_0)\rangle|^2.
\end{equation}
The LG modes $(m',n')$ are defined by
\begin{equation}
\label{eq:LGbeams}
\begin{split}
u_{mn}(r,\phi,z)= \sqrt{\frac{2n! }{\pi w_0^2\left(|m|+n\right)!}}  \left(\frac{\sqrt{2}r}{w(z)}\right)^{|m|} \exp\left(\frac{-r^2}{w^2(z)}\right) L^m_n\left(\frac{2r^2}{w^2(z)}\right)\\
\times \exp(im\phi)\exp\left(\frac{-ikr^2z}{2(z^2+z_R^2)}\right)\exp\left(i(2n+m+1)\tan^{-1}\left(\frac{z}{z_R}\right)\right)
 \end{split}
\end{equation}
where $z_R=\pi w_0^2/\lambda$ is the Rayleigh length with $w_0$ being the beam waist (at $z=0$), $L^m_n$ the associated Laguerre polynomials, $m\in [\infty,\infty]$ and $n\in[0,\infty]$ the azimuthal and radial mode index, respectively.\newline
In contrast to the mode decomposition procedure in \cite{RN365} the beam waists of incoming ($u^{LG}_{00}$) and outgoing beams ($u^{LG}_{m'n'}$) are not assumed equal, as there is no physical requirement for it \cite{Longman2017}. However, a varying beam size (as well as position of the beam that is neglected here) affects the mode decomposition result. In the calculations discussed here the output beam size is fixed to $\omega_0=$\SI{1.5}{\mu m} while the input beam size is varied. Fig.~\ref{fig:S1-MCEdecomposition} illustrates how the fundamental mode MCE changes upon variation of $\omega_0$ exhibiting a maximum at an optical beam waist ratio $\gamma=\omega_0/\omega_1$.
The maximum achievable MCE value is used as reference to compare between different mode converting optics. The optimum beam waist ratio increases with increasing topological charge of the SZP.

\section{Material property dependence of the mode conversion}
\begin{figure}
\begin{center}
\includegraphics{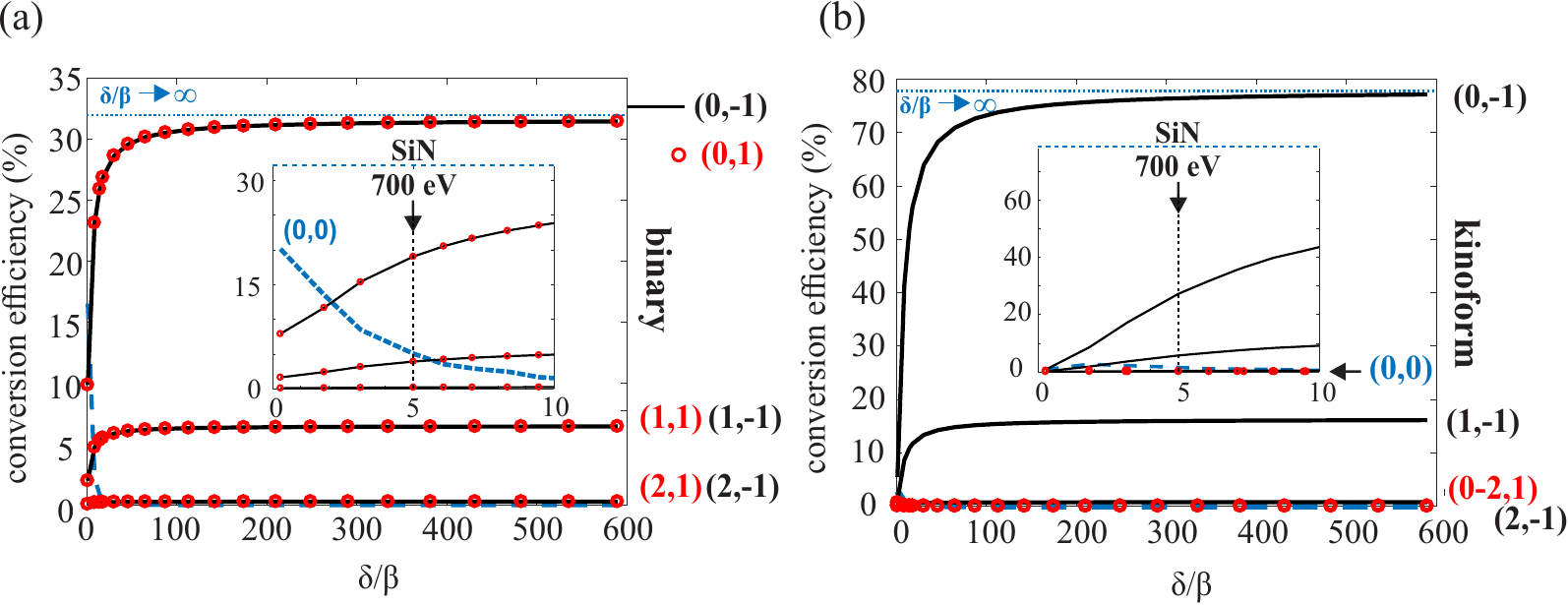}
\caption{\label{fig:S2-MCElimtis} \textbf{MCE of binary and kinoform in the limits of high and low absorption.} MCE into unconverted mode $(0,0)$ (dashed line), $(m=0-2,n=1)$ ($\circ$) and $(m=0-2,n=1)$ (solid line) for the (a) binary and (b) kinoform as a function of the dimensionless ratio $\delta / \beta$. The insets show magnifications of the low $\delta / \beta$ range. The dashed vertical line indicates $\delta /\beta$ for silicon nitride at \SI{700}{eV}. The SZP function and the MCE of the binary converges to the ideal binary for high absorption (low $\delta / \beta$). The kinoform converges to the ideal kinoform for decreasing absorption (high $\delta /\beta$). }
\end{center}
\end{figure}

Fig.~\ref{fig:S2-MCElimtis} depicts the MCE of the major mode contributions of both simulated binary and kinoform with material absorption in the limits of high $\delta/\beta \rightarrow 0$ and low absorption $\delta/\beta \rightarrow \infty $, where $\delta$ is the decrement of the real part and $\beta$ the imaginary part of the refractive index.
The graphs reveal that the MCEs of the binary converge to the binary in Fig.~2~(a) (main text) in the high absorption limit (see inset in Fig.~\ref{fig:S2-MCElimtis}~(a)) and the MCEs of the kinoform to Fig.~2~(b) in the low absorption limit (cf. Fig.~\ref{fig:S2-MCElimtis}~(b)).

\bibliography{SupplementalMaterial}